\documentclass{ws-ijmpd}
\usepackage[super,compress]{cite}
\begin{document}


%
%

\title{ASTROSAT: Some Key Science Prospects}

\author{Biswajit Paul\footnote{On behalf of the ASTROSAT Team}}

\address{Raman Research Institute\\
Sadashivanagar, C. V. Raman Avenue, Bangalore 560080, India\\
bpaul@rri.res.in}

\maketitle


\begin{abstract}
ASTROSAT is an astronomy satellite designed for simultaneous multi-wavelength studies in the Optical/UV and a broad X-ray energy range.
With four X-ray instruments and a pair of UV-Optical telescopes, ASTROSAT will provide unprecendented opportunity for simultaneous multi-wavelength observations, which is of immense value in study of highly variable sources, especially X-ray binaries and Active Galactic Nuclei.
The Large Area X-ray Proportional Counters (LAXPC) of ASTROSAT, which has the largest effective area in the hard X-ray band compared to all previous X-ray missions, will enable high time resolution X-ray measurements in the 2-80 keV band with moderate energy resolution.
Here we give a brief summary of the payload characteristics of ASTROSAT and discuss some of the main science topics that will be addressed with the LAXPC, and with simultaneous observations with the UVIT telescopes, with particular emphasis on X-ray binaries and compact objects.
The possibility of aiding gravitational wave experiments is also briefly mentioned.

\end{abstract}


\ccode{PACS numbers:}


\section{Introduction}

Almost all astrophysical objects have significant non-thermal processes that result in emission across a broad energy band. Even stars exhibit flares which are non-thermal processes. Multi-wavelength observations are therefore key to have complete understanding of a variety of astrophysical objects and astrophysical processes. X-ray binaries and Active Galactic Nuclei (AGN) are prime examples of objects with emission over a wide band. Often the most dominant part of the electromagnetic emission is within the optical to hard X-ray band. Moreover, these accretion powered sources show intensity variation over a wide range of timescales, from milliseconds which is the time scale for material to go around a low magnetic field neutron star, to years which is the time scale around supermassive black holes in the most powerful of AGNs. The X-ray variation can be periodic, quasi-periodic, random with flares, bursts and outbursts, and often a mixture of two or more of these patterns. The optical and UV emission also show intensity variations, though not quite as sharp and strong as X-rays. While in some sources one type of emission is known to be reprocessed into another type and therefore, the causal relation between the multiple components are understood to some extent, there are types of sources in which the source of variations, their causes and effects on the other components are not yet understood very well. Simultaneous multi-wavelength observations are therefore key to our understanding of the broadband emission mechanism and interrelation between multiple emission processes in X-ray binaries and AGNs. 

Simultaneous multi-wavelength X-ray and UV-Optical observations are however very difficult to carry out. Apart from the requirement of coordination between the space-based X-ray observatories and ground-based (or on a different space telescope like the HST) optical telescopes, the typical duty cycle for simultaneous multi-wavelength data from such observations is very low, owing to the different nature of Good Time Interval (GTI) of space and ground based observatories. Simultaneous X-ray and UV-Optical observations from the same satellite platform can mitigate many of these problems and yield rich dataset. ASTROSAT, with its three co-aligned X-ray instruments and two UV-Optical telescopes with a total of three bands is designed to bring in this new capability of quality simultaneous multi-wavelength observation capability. Two other space observatories, the XMM-Newton\cite{jansen2001} and Swift\cite{gehrels2004} have limited optical and UV capabilities along with very powerful X-ray telescopes and the new potential of single platform simultaneous multi-wavelength observations have been demonstrated to quite a large extent with these instruments.

In addition to the multi-wavelength aspect, the three co-aligned X-ray instruments will also provide very good capability for broadband X-ray measurements, with particular improvement in capabilities in the hard X-ray band of 15-80 keV. The Large Area X-ray Proportional Counter (LAXPC) instrument of ASTROSAT has an effective area that is in this range several times larger than any other X-ray instrument, which will enable certain unique measurements like Quasi Periodic Oscillations (QPO) in the hard X-ray band, pulse phase-resolved study of the Cyclotron Resonance Scattering Fearure (CRSF) in accretion-powered high magnetic field pulsars etc.

The ASTROSAT\cite{agrawal2006} is being developed under a large collaboration including several Indian organisations and some participation from the United Kingdom and Canada. Though the mission has been delayed, it is expected to be launched in 2013.
In the following sections we give a brief description of the instruments onboard ASTROSAT and discuss some of the main science topics that will be addressed with the LAXPC, and with simultaneous observations with the UVIT telescopes, with particular emphasis on X-ray binaries and compact objects.

\section{ASTROSAT}

There are five scientific payloads on ASTROSAT. Four of the instruments, including three in the X-ray band are co-aligned while the fifth instrument is an X-ray sky monitor. The sky monitor will scan about half of the sky in a given orientation of the spacecraft and as the spacecraft is pointed towards different sources, it will obtain all sky X-ray images to monitor and detect brightness variations in previously known bright X-ray sources and also to detect outbursts of new and known transient X-ray sources.

\subsection{Ultraviolet Imaging Telescope (UVIT)}
The UVIT consists of a pair of telescopes each with a primary aperture of 38 cm\cite{kumar2012}. One telescope is equipped with a Far-UV camera in the wavelength band of 1300-1800 $\rm \AA$ while the other telescope has a beam splitter and two cameras in the Near-UV range of 1800-3000 $\rm \AA$ and visible range of 3200-5300 $\rm \AA$. UVIT therefore provides three band simultaneous imaging over a field of view of 28 arcmin with an angular resolution of 1.8 arcsec. All the cameras are equipped with multiple narrow and broadband filters. UVIT is built to provide the highest angular resolution images in broad band UV among all space UV telescopes.

\subsection{Soft X-ray Telescope (SXT)}
The SXT is an imaging X-ray instrument with a grazing incidence X-ray telescope made with two sets of conical foil mirrors\cite{sagdeo2010}. At the focal plane is a CCD camera that has an energy resolution of about 200-250 eV at 6.4 keV while the telescope has an angular resolution in the range of 3-4 arcmin and field of view of 41 arcmin. As a supporting instrument, the SXT broadens the energy range of the X-ray instruments of ASTROSAT at the lower end to about 0.35 keV. The SXT will also be useful to determine positions of new transient X-ray sources detected with the sky monitor to within a fraction of an arcmin depending on the brightness.

\subsection{Cadmium-Zink-Telluride Imager (CZTI)}
The CZTI is a coded aperture mask instrument made of pixelated cadmium-zinc-telluride (CZT) detectors of area 1024 cm$^2$ in the image plane and a coded mask over it at a height of about 70 cm. The CZTI has a field of view of 6x6 square degree in the low energy band and a wider field of view of 17x17 square degree above 100 keV. It is designed to provide an angular resolution of eight arcmin and an energy resolution of $\sim2\%$ in the energy band of 10-150 keV. Due to a coded aperture mask design, the sensitivity of the CZTI depends on the presence of other bright sources in the field of view and it is different in crowded fields.

\subsection{Large Area X-ray Proportional Counter (LAXPC)}
The LAXPC is a set of three xenon gas proportional counter detectors, with a total geometrical area of 10,800 cm$^2$ and open area in excess of 6,000 cm$^2$\cite{paul2009}. It is designed to work in the energy band of 2-80 keV with an effective area in excess of 6000 cm$^2$ over a large energy range of 3-20 keV and in excess of 2,400 cm$^2$ upto 80 keV (Figure 1). With such large effective area over a wide energy band, the LAXPC will be the workhorse of ASTROSAT for X-ray timing. Along with the SXT in the lower energy side and the CZT in the higher energy side, it will also be very useful for broad band X-ray spectroscopy. The energy resolution of LAXPC is moderate, in the range of 14\%-18\% in the entire energy band of LAXPC. It is designed to have a timing resolution and absolute timing accuracy of 10 micro seconds, suitable for study of rapid variability in X-ray binaries.

\begin{figure}[pt]
\centerline{\psfig{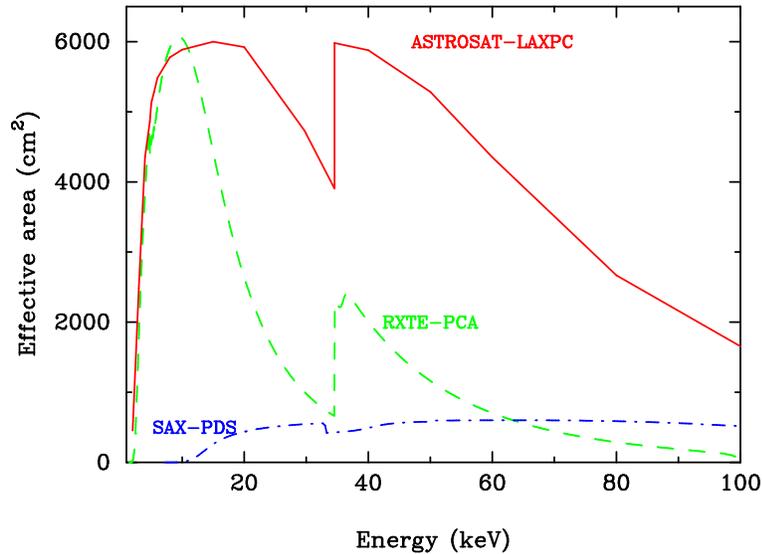}}
\vspace*{8pt}
\caption{The effective area curve of ASTROSAT-LAXPC is shown here along with the same of RXTE-PCA and SAX-PDS.
\label{f1}}
\end{figure}

\subsection{Scanning X-ray Sky Monitor (SSM)}
The SSM consists of three one dimensional position sensitive proportional counters with one dimensional coded aperture mask\cite{seetha2006}. The instantaneous sky coverage of each detector-mask set is in the form of a long strip and the sky coverage is obtained by rotating the detectors around an axis perpendicular to the viewing axis of the four co-aligned instruments of ASTROSAT. The SSM works in the energy band of 2-12 keV and has a sky coverage of about 50\% for a given pointing of the spacecraft.

\section{Key Features of ASTROSAT}

The key features of ASTROSAT are:
\begin{itemize}
\item X-ray and UV-Optical instruments on the same satellite.
\item High angular resolution, simultaneous three band imaging with the UVIT.
\item Wide energy band in the X-rays, suitable for broad band X-ray spectroscopy.
\item Large effective area in the hard X-rays, suitable for hard X-ray timing studies.
\item All sky monitoring from the same satellite, opportunity for quick follow-up observations.
\item Near equatorial orbit of the satellite resulting in low and stable X-ray background, suitable for accurate estimation of background rate.
\end{itemize}

\subsection{Comparison with contemporary space missions with similar capabilities}

Other space observatories with multi-wavelength capabilities are XMM-Newton\cite{jansen2001} and Swift\cite{gehrels2004}. Both these observatories have major X-ray instruments: the XMM-Newton observatory with high throughput telescopes and multipurpose focal plane instruments including X-ray imagers and Reflection Grating Spectroscopy (RGS) and the Swift observatory with wide field Burst Alert Telescope (BAT) and X-ray Telescope (XRT) for narrow field imaging and spectroscopy. However, the optical and the UV telescopes on these missions have limited sensitivity.
The ultraviolet imaging observatory Galex\cite{martin2005} has sensitivity comparable to the UVIT. However, the UVIT, with an angular resolution of 1.8 arcsec compared to about 5 arcsec of Galex has advantage in crowded fields and simultaneous three band imaging is another advantage with the UVIT.

In the X-ray band alone, there are two aspects of ASTROSAT that can be compared with the other space observatories: broad band spectroscopy and fast timing. The widest spectral coverage is available in the 5th Japanese X-ray astronomy mission Suzaku\cite{mitsuda2007}. The hard X-ray instrument of Suzaku, having a small area but very low background is more suitable for spectroscopy of faint objects than timing. In comparison, the ASTROSAT-LAXPC, with large collective area and relatively higher background is more suitable for timing studies and fast spectroscopy for bright sources. Another wide band X-ray observatory was the Dutch-Italian mission Beppo-SAX\cite{frontera1997} and similar comparison is valid with the ASTROSAT-LAXPC.
In X-ray timing studies, the major milestones are EXOSAT\cite{turner1981}, GINGA\cite{turner1989} and the RXTE\cite{jahoda1996} . The RXTE, with an effective area of 6,000 cm$^2$ and a high time resolution has pioneered fast timing studies. The ASTROSAT-LAXPC, being the only timing instrument in space in its timeframe has very high potential for all types of fast timing studies. Of particular benefit will be the several times larger effective area of LAXPC compared to the RXTE-PCA in the hard X-ray band (Figure 1).

\section{Key Science Topics with LAXPC}

The LAXPC instrument will provide some unique scientific prospects for the following reasons:

\begin{itemize}
\item After the very successful operation of the RXTE-PCA during 1996-2012, LAXPC is the only X-ray timing instrument in its time frame: Some of the new outbursts of known transient sources and new transients to be discovered with the SSM or other X-ray monitors will be followed up with the LAXPC. 
\item The LAXPC has a wider energy band compared to all previous large area instruments: All types of timing features studied with the previous observatories will now be investigated over wider energy band leading to better/tighter constraints for models.
\item The effective area in the hard X-ray band is several times larger than all other hard X-ray detectors: Physical processes that dominate in the hard X-ray band will be better probed with the LAXPC, especially processes with some temporal characteristics.
\item There are other soft and hard X-ray detectors extending the energy band for broad band spectroscopic studies: This will lead to better constraints on the continuum spectral model components.
\item A near equatorial orbit will give low and stable background rate: This will allow observations of fainter sources compared to what is possible with an instrument with large and variable background.
\item Opportunity for simultaneous multi-savelength observations: This is an unique and almost new feature for a space observatory. 
\end{itemize} 

Here we discuss some of the scientific issues that will be pursued well with the LAXPC. Some of these are to be pursued together with the other instruments of
ASTROSAT.

\subsection{Broadband X-ray spectroscopy}

Many astrophysical objects have complex and multiple physical processes ongoing simultaneously. Among the X-ray sources, there are very few types of objects which produce X-rays by a single emission mechanism. The relative importance of the difference processes and their interplay, often manifests in a wide energy band and measuring correlation of different spectral components and constraining their relative strengths is crucial. The multiple, complex, and interrelated high energy processes are simultanelusly observed very well by broad band spectroscopy in the binary black holes\cite{frontera2001} and accretion powered X-ray pulsars\cite{dalfiume2000}.
The broadband spectral capability
of ASTROSAT will therefore be very suitable for detailed investigations of
a wide range of X-ray sources.

\subsection{Cyclotron Resonance Scattering Feature (CRSF)}
The accreting, high magnetic field neutron stars called binary X-ray pulsars
show strong hard X-ray emission\cite{nagase1989}, with a power-law spectral shape and a photon
index of about 1.0. The X-ray spectrum usually has a high energy cutoff at around 20 keV above which the spectrum shows exponential fall with an e-folding energy of about 15 keV. The hard X-ray spectrum is understood to be produced by inverse Compton scattering of soft X-ray photons by energetic electrons in the accretion column/slab on the magnetic poles of the neutron star. An interesting feature of the X-ray spectrum that has been observed in about 20 such sources is that there are broad absorption features\cite{coburn2006}, which are understood to be due to cyclotron resonance scattering of photons by elctrons in presence of the strong surface magnetic field (a few times 10$^{12}$ Gauss) of the neutron star. The CRSF often shows harmonics and several harmonics have been simultaneously detected in the source 4U 0115+63\cite{santangelo1999}, which of course requires broad band X-ray detectors.

\begin{figure}[pt]
\centerline{\psfig{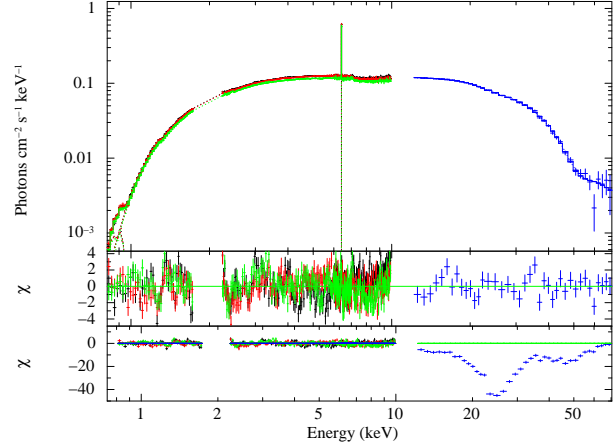}}
\vspace*{8pt}
\caption{The broad band deconvolved X-ray spectrum of Vela X-1 observed with the Suzaku observatory is shown here with the best fit model spectrum including a high energy cut-off power law along with partial covering absorption and cyclotron resonance scattering feature at 26 keV and its harmonic. Residuals to the best fitted spectra are shown in the middle panel and residuals without including the CRSF are shown in the bottom panel. Credit: {\it Chandreyee Maitra}.
\label{f2}}
\end{figure}

The CRSF is a measure of the strength of the magnetic field of the neutron star (CRSF at 12 keV corresponds to B $\sim$10$^{12}$ Gauss) and is the only direct method. Moreover, a study of the CRSF provides various ways to understand the accretion phenomena onto high magnetic field neutron stars and complex physical processes and geometrical aspects of the accretion column/slab. The transient X-ray pulsars provide us a way to study the dependence of the processes on the mass accretion rate, and a study of the CRSF energy against the X-ray luminosity has shown interesting results: anticorrelation between the CRSF energy and X-ray luminosity in some sources and positive correlation between the two in some other sources\cite{nakajima2006,klochkov2012,becker2012}.

\begin{figure}[pt]
\centerline{\psfig{file=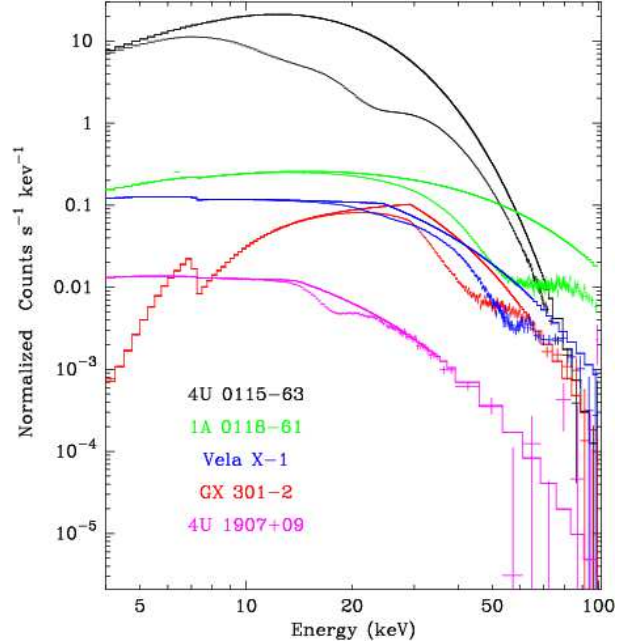,angle=-90,width=8cm}}
\vspace*{8pt}
\caption{Simulated X-ray spectra of several pulsars as would be detected with the ASTROSAT-LAXPC. Credit: {\it Chandreyee Maitra}.
\label{f1}}
\end{figure}

The pulsars are interesting in another way as the rotation allows us to look at the neutron star and the magnetic field structure from different angles, a novelty in astronomy. Thus pulse phase resolved measurements of the CRSF is a very useful way to investigate the accretion process onto the highly magnetized neutron stars. It has barely been possible to study pulse phase dependence of the CRSF in a few of the brighest pulsars\cite{suchy2012, maitra2012a, maitra2012b}, thanks to the broad band spectral coverage with the Suzaku observatory. The broad band X-ray spectrum including the CRSF feature and its harmonic detected in the X-ray pulsar Vela X-1 is shown in Figure 2. From the same observation with the Suzaku observatory, strong pulse phase dependence of the CRSF parameters have been found including a variation in the ratio of energies of the fundamental and harmonics of the CRSF\cite{maitra2012b}.

The pulsar cyclotron line studies, both the luminosity dependendence and the pulse phase dependence can be carried out very efficiently with the ASTROSAT-LAXPC owing to its large effective area in the hard X-rays for these very bright sources. A simulation of the CRSF, as would be detected in observations of several bright X-ray pulsars are shown in Figure 3.

\begin{figure}[pt]
\centerline{\psfig{file=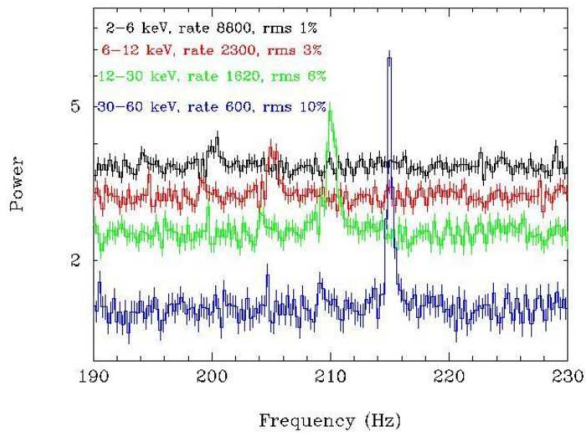,angle=00,width=8cm}}
\vspace*{8pt}
\caption{Simulation of energy dependent QPO detections with the ASTROSAT-LAXPC. The incident X-ray spectrum and energy dependence of the QPO rms are taken for real sources.
\label{f1}}
\end{figure}

\subsection{Quasi Periodic Oscillations (QPO) in the hard X-rays}

The QPO phenomena in X-ray binaries has been known for almost three decades\cite{hasinger1986}. The QPOs are found in all types of X-ray binaries and they appear in different varities in sources with different types of compact stars (black holes, low magnetic field neutron stars, high magnetic field neutron stars) and even in one type of source, there are different distinct types of QPOs. The most discussed, and perhaps scientifically most valuable are the kiloHertz QPOs in low mass X-ray binaries\cite{vanderklis1997} with low magnetic field neutron stars and the rarely occuring high frequency QPOs in black hole X-ray binaries\cite{belloni2012}. QPOs have now also been observed in intermediate mass black hole systems and some AGNs\cite{dewangan2006}.
However, inspite of being very rich in observable features, QPOs are one of the least understood aspect of X-ray binaries. The LAXPC provides a new way of investigate the QPOs, i.e. in the hard X-ray band. The QPOs show a trend of increasing rms with energy, at least upto 25 keV. In spite of falling number of photons with energy in all X-ray binaries, the large effective area of the LAXPC in hard X-rays, combined with the large rms variation of the source photons in hard X-rays give a high signal to noise ratio for QPO detection at hard X-rays. In Figure 4, a simulation shows the QPO features in a black hole X-ray binary as would be detected with LAXPC in different energy bands.

\begin{figure}[pt]
\centerline{\psfig{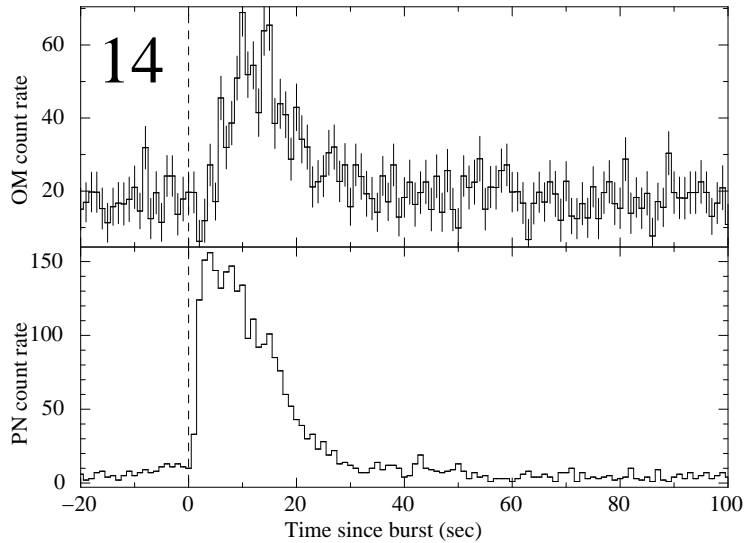}}
\vspace*{8pt}
\caption{A thermonuclear X-ray burst and reprocessed optical emission in EXO 0748-676\cite{paul2012}, detected with the XMM-Newton observatory.
\label{f1}}
\end{figure}

\subsection{Thermonuclear bursts and reprocessing of X-rays}

Some of the accreting low magnetic field neutron stars show thermonuclear X-ray bursts. Measurement of the luminosity and X-ray spectral evolution during the bursts is a very useful way to invesigate the mass-radius relation of neutron stars, a very important topic in astrophysics. Kilohertz oscillations have also been detected during the thermonuclear bursts, the study of which gives clues to the neutron star equation of state and also to the process of burst ignition and propagation\cite{bhattacharyya2005}. The thermonuclear bursts are also very useful to investigate properties of the X-ray binaries and the physics of reprocessing of X-rays in plasma\cite{hynes2006}. In the low mass X-ray binaries the companion star is often not dominant in the optical. The UV and optical emission is partly from energy dissipation in the accretion disk, and partly from reprocessing of X-rays from the compact star. The thermonuclear X-ray bursts provide a unique time varying input, and simultaneous measurement of the X-ray bursts and the reprocessed burst in UV and optical is a very powerful tool to understand the X-ray reprocessing mechanisms and parameters of the binary. Known orbital parameters of some X-ray binaries, and opportunity to study some sources in different orbital phases are additional advantages.

However, the difficulties in carrying out simultaneous multiwavelength observations and relative rarity of the thermonuclear bursts (a few hours interval on average) has limited the growth of such studies. The LAXPC is well suited to carry out different types of studies of the LMXBs. Along with measurements of the high frequency periodic and aperiodic timing features, the LAXPC will be very useful to tightly constrain the bolometric luminosity, especially near the peak of the bursts when a temperature in excess of 2 keV is reached. The simultaneous UV-Optical data will also be very useful for study of the X-ray reprocessing. Detection of bursts from the X-ray source EXO 0748-676 with the XMM-Newton optical monitor\cite{paul2012} shown in Figure 5, demonstrates the suitability of ASTROSAT for such observations, but with three bands in the UV-Optical.

\begin{figure}[pt]
\centerline{\psfig{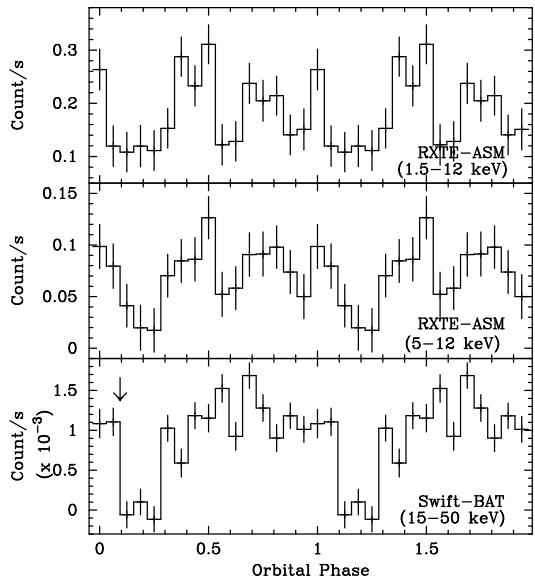}}
\vspace*{8pt}
\caption{X-ray light curve of the SFXT IGR J16479-4514 showing the orbital modulation with an orbital period of 3.32 days\cite{jain2009}. The eclipse, seen clearly in the bottom panel lasts for about 25\% of the orbital period.
\label{f1}}
\end{figure}

\subsection{Supergiant Fast X-ray Transients (SFXT)}
In the last five years or so a new class of binary X-ray sources, named
SFXTs, have been of great interest\cite{sidoli2011}. As the name suggests,
in these X-ray binaries the companion star is a supergiant star and these
sources show fast X-ray transient phenomena, ranging from a few minutes to
several hours. Models proposed to explain the SFXT phenomena include gated
accretion onto neutron stars with strong magnetic field, accretion from
dense clumps in the winds of the companion star etc. During the non-burst
periods, the SFXTs have an X-ray intensity that is about two orders of magnitude
lower than the typical high mass X-ray binaries with supergiant companions
like Vela X-1. The low average luminosity is understood to be due to the
wide orbits and corresponding large orbital periods in these binaries. 
However, some exceptions like IGR J16479-4514 which has a small orbital
period\cite{jain2009} of only 3.2 days (Figure 6) raises doubts about the
proposed scenarios. Though the compact objects in most of the SFXTs are expected
to be high
magnetic field neutron stars, pulsations have been detected in only a few of
these sources. Observations with the LAXPC will be useful to make careful
and sensitive searches for pulsations in more of these objects. If found,
the pulsations will be very useful to measure the binary parameters of the
SFXTs.

\subsection{Gravitational Wave (GW) radiation from accreting neutron stars in X-ray binaries}

accreting neutron stars are one of the different types of sources of
GW emission that could be detected with the ground based
GW detectors. For example, an accretion mound on a rapidly
rotating neutron star can be a reasonably strong source and candidate
for detection. However, to search for GW at a few hundred Hz from a long
data base of a few months or years, it is essential to know the spin and
orbital parameters of the neutron star and its changes over the entire
period. Otherwise, the parameters space for search is very large\cite{watts2009}.
Naturally, the sources with high accretion rate are likely to be stronger
sources of GW but the spin and orbital parameters are least constrained for such
sources. The sources with best constrained solutions for spin and orbital
parameters, the millisecond accreting pulsars, on the other hand have low
long term averge accretion rate. If LAXPC observations with improved sensitivity
over RXTE-PCA leads to measurement of the spin and orbital solution for some of
the higher accretion rate LMXBs (see Jain \& Paul 2011 for example), this could
be a useful input for search of GW from such sources.


\end{document}